\documentclass[american,english]{IEEEtran}
\usepackage[T1]{fontenc}
\usepackage[latin9]{inputenc}
\usepackage{color}
\usepackage{mathtools}
\usepackage{amsmath}
\usepackage{amssymb}
\usepackage{cancel}
\usepackage{mathdots}
\usepackage{stmaryrd}
\usepackage{stackrel}
\usepackage{graphicx}
\PassOptionsToPackage{version=3}{mhchem}
\usepackage{mhchem}
\usepackage{esint}

\makeatletter
\usepackage{colortbl}
\usepackage{cite}

\usepackage{flushend}
\usepackage{stfloats}
\usepackage[margin=8pt,font=footnotesize,labelfont=md,labelsep=colon]{caption}

\usepackage{multicol}
\usepackage{float}
\usepackage{lipsum}
\usepackage[figurename=Fig.]{caption}

\makeatother

\usepackage{babel}
\begin{document}
\title{Physical Layer Security in Vehicular Networks with Reconfigurable
Intelligent Surfaces}
\author{\textcolor{black}{\normalsize{}Abubakar U. Makarfi}\textit{\textcolor{black}{\normalsize{}$^{1}$,}}\textcolor{black}{\normalsize{}
Khaled M. Rabie}\textit{\textcolor{black}{\normalsize{}$^{2}$}}\textcolor{black}{\normalsize{},
Omprakash Kaiwartya}\textit{\textcolor{black}{\normalsize{}$^{3}$}}\textcolor{black}{\normalsize{},
Xingwang Li}\textit{\textcolor{black}{\normalsize{}$^{4}$, }}\textcolor{black}{\normalsize{}Rupak
Kharel}\textit{\textcolor{black}{\normalsize{}$^{1}$}}\textcolor{black}{\normalsize{}}\\
\textcolor{black}{\normalsize{}$^{1}$Department of Computing and
Mathematics, Manchester Metropolitan University, UK}\\
\textit{\textcolor{black}{\normalsize{}$^{2}$}}\textcolor{black}{\normalsize{}Department
of Engineering, Manchester Metropolitan University, UK}\\
\textit{\textcolor{black}{\normalsize{}$^{3}$}}\textcolor{black}{\normalsize{}School
of Science and Technology, Nottingham Trent University, UK }\\
\textit{\textcolor{black}{\normalsize{}$^{4}$}}\textcolor{black}{\normalsize{}School
of Physics and Electronic Information Engineering, Henan Polytechnic
University, China }\\
\textcolor{black}{\normalsize{}Emails:\{a.makarfi, r.kharel, k.rabie\}@mmu.ac.uk;
omprakash.kaiwartya@ntu.ac.uk; lixingwang@hpu.edu.cn. }}

\maketitle
\selectlanguage{american}%
\textcolor{black}{\thispagestyle{empty}}
\selectlanguage{english}%
\begin{abstract}
This paper studies the physical layer security (PLS) of a vehicular
network employing a reconfigurable intelligent surface (RIS). RIS
technologies are emerging as an important paradigm for the realisation
of smart radio environments, where large numbers of small, low-cost
and passive elements, reflect the incident signal with an adjustable
phase shift without requiring a dedicated energy source. Inspired
by the promising potential of RIS-based transmission, we investigate
two vehicular network system models: One with vehicle-to-vehicle communication
with the source employing a RIS-based access point, and the other
model in the form of a vehicular adhoc network (VANET), with a RIS-based
relay deployed on a building. Both models assume the presence of an
eavesdropper, to investigate the average secrecy capacity of the considered
systems. Monte-Carlo simulations are provided throughout to validate
the results. The results show that performance of the system in terms
of the secrecy capacity is affected by the location of the RIS-relay
and the number of RIS cells. The effect of other system parameters
such as source power and eavesdropper distances are also studied.
\end{abstract}


\section{Introduction}

Recent research in beyond 5G technologies has brought
about new communication paradigms, especially at the physical layer.
One of such emerging paradigms is the concept of ``smart radio environments'',
enabled by technologies to control the propagation environment in
order to improve signal quality and coverage, such
as reflector-arrays/intelligent walls and reconfigurable intelligent
surfaces (RIS) \cite{Renzo2019,Liaskos_metasurface}.
RISs are man-made surfaces of electromagnetic material that are electronically
controlled with integrated electronics and have unique wireless communication
capabilities \cite{Basar2019WirelessCT}. 

RIS-based transmission concepts have been shown to be completely different
from existing MIMO, beamforming and amplify-and-forward/decode-and-forward
relaying paradigms, where the large number of small, low-cost and
passive elements on a RIS only reflect the incident signal with an
adjustable phase shift without requiring a dedicated energy source
for RF processing or retransmission \cite{Basar_xmsn_LIS}. Applications
of RIS have recently been investigated with respect to signal-to-noise
ratio (SNR) maximisation \cite{Basar_xmsn_LIS}, improving signal
coverage \cite{subrt}, improving massive MIMO systems \cite{MIMO_LIS},
beamforming optimisation \cite{Wu_beam_opt,Wu_beamform,7510962},
as well as multi-user networks\cite{LIS_multi_user}. 

On the other hand, physical layer security (PLS) in vehicular networks
is important due to rapid advancements towards autonomous vehicles
and smart/cognitive transportation networks to minimise the risk from
compromise. Significant research efforts have been expended in studying
vehicular networks with \cite{psl_v2v} or without PLS \cite{relay_dbl_ray,fuzzyVCPS,geometryGPSoutage_ompra,measuresVCPS}.
Given that PLS employs the inherent characteristics of the propagation
channels, such as interference, fading and noise to realise keyless
secure transmission through signal processing approaches {[}2{]},
then RIS-based systems are well positioned for such applications.
Initial results have been reported for PLS studies for systems employing
RIS technologies \cite{secure_IRS,Chen2019IntelligentRS}. However,
no such studies have been conducted for vehicular adhoc networks (VANETs).

From the aforementioned, this study presents the
following contributions. We study the PLS of two possible vehicular
network models, by analysing and deriving expressions for the average
secrecy capacity. The first model considers a vehicle-to-vehicle (V2V)
network with the source vehicle employing a RIS-based access point
(AP) for transmission. In the second model, we consider a VANET with
a source station transmission via a RIS-based relay. Such a RIS-relay
could be deployed on a building as part of a smart infrastructure
within a city. To the best of our knowledge, this is the first analysis
of a RIS-based technology employed within a VANET. Moreover, the distances
of the legitimate and eavesdropper nodes are taking into account along
with realistic fading scenarios considered for the base stations and
the mobile nodes. Monte Carlo simulations are provided throughout
to verify the accuracy of our analysis. The results show that the
performance of the system in terms of the secrecy capacity is improved
with the use of the RIS. Furthermore, the effect of the system parameters
such as source power, eavesdropper distance and number of RIS cells
on the system performance are investigated.

The paper is organised as follows. In Sections \ref{sec:RIS AP}
and \ref{sec:RIS-relay}, we describe the two system models under
study and analyse the secrecy performance by deriving accurate analytical
expressions for efficient computation of the secrecy capacity of the
networks. Finally, in Sections \ref{sec:Results} and \ref{sec:Conclusions},
we present the results with discussions and outline the main conclusions,
respectively.

\section{V2V with RIS as Access Point\label{sec:RIS AP}}

\begin{figure}[th]
\begin{centering}
\includegraphics[scale=0.4]{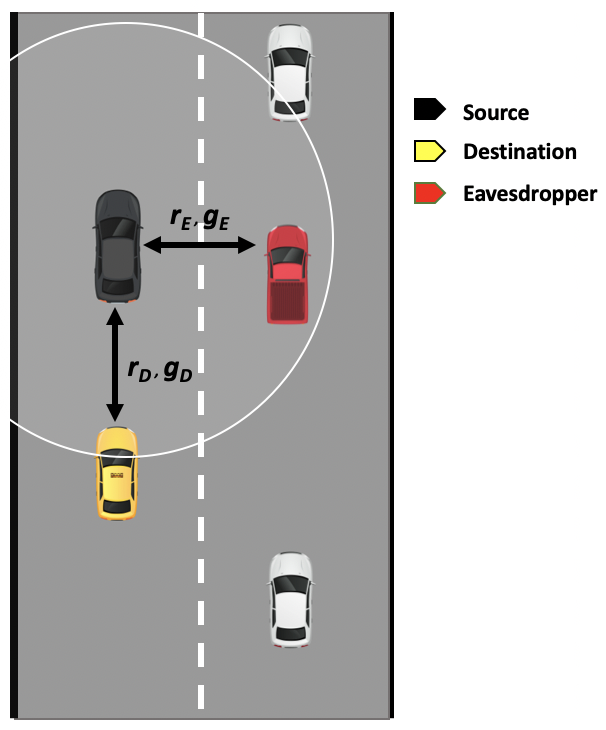}
\par\end{centering}
\caption{System model for V2V scenario with the source vehicle using RIS as
AP.\label{fig:sys-mod}}
\end{figure}
\begin{figure}[th]
\begin{centering}
\includegraphics[scale=0.35]{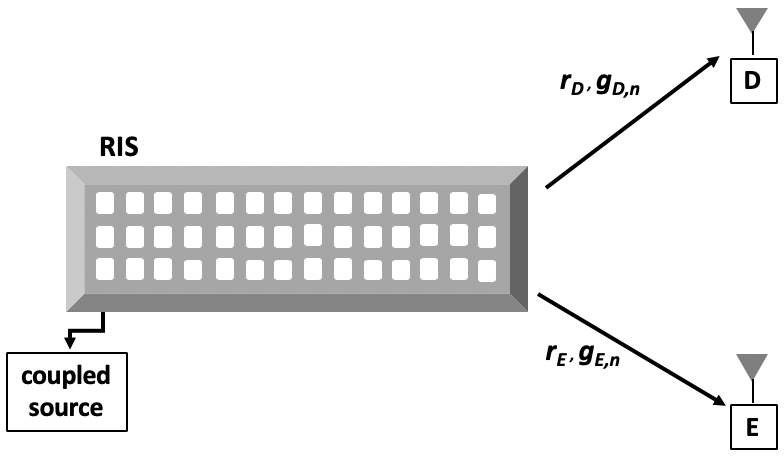}
\par\end{centering}
\caption{Vehicle RIS configuration as an AP.\label{fig:sys-mod-2}}
\end{figure}

In this section, we describe the V2V network with RIS as AP and derive
expressions for the average secrecy capacity of the system.

\subsection{System Description}

We consider a system of vehicles operating in a network as shown in
Fig. \ref{fig:sys-mod}. We assume a classic Wyner's wiretap model
in our analysis \cite{Lei_sec_cap}, such that an information source
vehicle ($S$), sends confidential information to a destination vehicle
($D$), while a passive eavesdropper\footnote{Passive eavesdropper in the sense that the node only intercepts the
information, but makes no attempt to actively disrupt, such as through
jamming.} vehicle ($E$) attempts to receive and decode the confidential information.
The vehicles $D$ and $E$ are known to lie within a certain radius
from $S,$ the precise relative distances of the V2V links are unknown
during transmission, which is a realistic assumption for a network
of this nature \cite{pls_eve_uncertain,psl_v2v}. Moreover, $S$ is
assumed to employ a RIS-based scheme in the form of an AP to communicate
over the network\footnote{Initial proposal and results for such a configuration of intelligent
surfaces were reported in \cite{Basar_xmsn_LIS}.}. As shown in the block diagram of Fig. \ref{fig:sys-mod-2}, the
RIS can be connected over a wired link or optical fiber for direct
transmission from $S$, and can support transmission without RF processing.
For the system considered, we assume an intelligent AP with the RIS
having knowledge of channel phase terms, such that the RIS-induced
phases can be adjusted to maximise the received SNR through appropriate
phase cancellations and proper alignment of reflected signals from
the intelligent surface.

The received signals at $D$ and $E$ are respectively represented
as
\begin{align}
y_{D} & =\left[\sum_{n=1}^{N}h_{D,n}e^{-j\phi_{n}}\right]x+w_{D},\label{eq:yD}\\
y_{E} & =\left[\sum_{n=1}^{N}h_{E,n}e^{-j\phi_{n}}\right]x+w_{E},\label{eq:yE}
\end{align}
where $x$ represents the transmitted signal by $S$ with power $P_{s}$,
while the terms $w_{D}$ and $w_{E}$ are the respective additive
white Gaussian noise (AWGN) at $D$ and $E$. Without loss of generality,
we denote the power spectral density of the AWGN as $N_{0}$ and equal
at both links. The terms $h_{i,n}=\sqrt{g_{i,n}r_{i}^{-\beta},}\quad i\in\left\{ D,E\right\} $,
is the channel coefficient from $S$ to the receiving vehicles $D$
and $E,$ where $r_{i}$ is the V2V link distance, $\beta$ is the
path-loss exponent and $g_{i,n}$ is the channel gain from the RIS
to the receiver, following independent double Rayleigh fading \cite{psl_v2v}.
The term $\phi_{n}$ is the reconfigurable phase induced by the $n$th
reflector of the RIS, which through phase matching, the SNR of the
received signals can be maximised\footnote{For the sake of brevity, the reader is referred to \cite{Basar_xmsn_LIS},
for details of phase cancellation techniques.}.

Based on (\ref{eq:yD}) and (\ref{eq:yE}), the instanstaneous SNRs
at $D$ and $E$ are given by
\begin{equation}
\gamma_{D}=\frac{\sum_{n=1}^{N}P_{s}\mid h_{D,n}\mid^{2}}{N_{0}},\label{eq:sinr-d}
\end{equation}
and

\begin{equation}
\gamma_{E}=\frac{\sum_{n=1}^{N}P_{s}\mid h_{E,n}\mid^{2}}{N_{0}}.\label{eq:sinr-e}
\end{equation}

\subsection{Average Secrecy Capacity Analysis \label{sec:Perf-anal}}

In this section, we derive analytical expressions for the secrecy
capacity of the system. The maximum achievable secrecy capacity is
defined by \cite{bloch_cap}
\begin{equation}
C_{s}=\max\left\{ C_{D}-C_{E},0\right\} ,\label{eq:cs-defined}
\end{equation}
where $C_{D}=\log_{2}\left(1+\gamma_{D}\right)$ and $C_{E}=\log_{2}\left(1+\gamma_{E}\right)$
are the instantaneous capacities of the main and eavesdropping links,
respectively. The secrecy capacity in (\ref{eq:cs-defined}) can therefore
be expressed as \cite{bloch_cap}
\begin{equation}
C_{s}=\begin{cases}
\log_{2}\left(1+\gamma_{D}\right)-\log_{2}\left(1+\gamma_{E}\right), & \gamma_{D}>\gamma_{E},\\
0, & \gamma_{D}<\gamma_{E}.
\end{cases}\label{eq:cs-defined-2}
\end{equation}

The average secrecy capacity $\overline{C_{s}}$ is given by \cite{Osamah18GlobecomSC}
\begin{align}
\overline{C_{s}} & =\mathbb{E}\left[C_{s}\left(\gamma_{D},\gamma_{E}\right)\right]\nonumber \\
 & =\stackrel[0]{\infty}{\int}\stackrel[0]{\infty}{\int}C_{s}\left(\gamma_{D},\gamma_{E}\right)f\left(\gamma_{D},\gamma_{E}\right)\textrm{d}\gamma_{D}\textrm{d}\gamma_{E},\label{eq:av-sec-cap-1}
\end{align}
where $\mathbb{E}\left[\cdot\right]$ is the expectation operator
and $f\left(\gamma_{D},\gamma_{E}\right)$ is the joint PDF of $\gamma_{D}$
and $\gamma_{E}$. In order to simplify the analysis, we express the
logarithmic function in (\ref{eq:cs-defined}) in an alternate form.
Recalling the identity \cite[Eq. (6)]{hamdi_cap_mrc}
\begin{equation}
\textrm{ln}\left(1+\zeta\right)=\stackrel[0]{\infty}{\int}\frac{1}{s}\left(1-e^{-\zeta s}\right)e^{-s}\textrm{d}s,\label{eq:log-id-1}
\end{equation}
and by substituting $\zeta=\gamma_{D}$ in (\ref{eq:log-id-1}), we
can express the instantaneous capacity of the main link as
\begin{equation}
\overline{C}_{D}=\frac{1}{\textrm{ln}\left(2\right)}\stackrel[0]{\infty}{\int}\frac{1}{z}\left(1-\mathcal{M}_{D}\left(z\right)\right)e^{-z}\textrm{d}z,\label{eq: log-id-2}
\end{equation}
where $\mathcal{M}_{D}\left(z\right)=\mathbb{E}\left[e^{-z\frac{P_{s}r_{D}^{-\beta}}{N_{0}}\sum_{n=1}^{N}g_{D,n}}\right]$
is the MGF of the SNR at $D$. 

Next, we compute the MGF $\mathcal{M}_{D}\left(z\right)$, defined
by
\begin{align}
\mathcal{M}_{D}\left(z\right)= & \mathbb{E}\left[e^{-z\frac{P_{s}r_{D}^{-\beta}}{N_{0}}\sum_{n=1}^{N}g_{D,n}}\right]\nonumber \\
= & \prod_{n=1}^{N}\mathbb{E}\left[e^{-z\frac{P_{s}r_{D}^{-\beta}}{N_{0}}g_{D,n}}\right]\nonumber \\
= & \prod_{n=1}^{N}\stackrel[g]{}{\int}e^{-z\xi_{D}g_{D,n}}f_{g_{D}}(g)\textrm{d}g_{D},\label{eq:mgf-D-1}
\end{align}
then from the generalized cascaded Rayleigh distribution, we can obtain
the PDF of the double Rayleigh channel for $n=2$ in \cite[Eq. (8)]{cascaded_ray}
as $f\left(g\right)=\textrm{G}_{0,2}^{2,0}\left(\frac{1}{4}g^{2}\Biggl|\negthickspace\begin{array}{c}
-\\
\frac{1}{2},\negthickspace\frac{1}{2}
\end{array}\negthickspace\right).$ By invoking \cite[Eq. (9.34.3)]{book2}, we can express the PDF by
re-writing the Meijer G-function in an alternate form. Thus, we get
\begin{equation}
f\left(g\right)=\textrm{G}_{0,2}^{2,0}\left(\frac{1}{4}g^{2}\Biggl|\negthickspace\begin{array}{c}
-\\
\frac{1}{2},\negthickspace\frac{1}{2}
\end{array}\negthickspace\right)=gK_{0}\left(g\right).\label{eq:meijer-bessel}
\end{equation}

Using (\ref{eq:meijer-bessel}) and \cite[Eq. (6.621.3)]{book2} along
with some basic algebraic manipulations, we can obtain the desired
result as 
\begin{equation}
\mathcal{M}_{D}\left(z\right)=\prod_{n=1}^{N}\frac{4}{3(1+z\frac{P_{s}r_{D}^{-\beta}}{N_{0}})^{2}}{}_{\phantom{}2}F_{1}\left(2,\frac{1}{2},\frac{5}{2},\frac{z\frac{P_{s}r_{D}^{-\beta}}{N_{0}}-1}{z\frac{P_{s}r_{D}^{-\beta}}{N_{0}}+1}\right).\label{eq:mgf-D-final}
\end{equation}

\textcolor{black}{Using similar analysis, the average capacity of
the eavesdropper link can be represented as}
\begin{equation}
\overline{C}_{E}=\frac{1}{\textrm{ln}\left(2\right)}\stackrel[0]{\infty}{\int}\frac{1}{z}\left(1-\mathcal{M}_{E}\left(z\right)\right)e^{-z}\textrm{d}z,\label{eq:av-CE-defined}
\end{equation}
where the MGF $\mathcal{M}_{E}\left(z\right)=\mathbb{E}\left[e^{-z\frac{P_{s}r_{E}^{-\beta}}{N_{0}}\sum_{n=1}^{N}g_{E}}\right]$
and can be similarly evaluated as 
\begin{equation}
\mathcal{M}_{E}\left(z\right)=\prod_{n=1}^{N}\frac{4}{3(1+z\frac{P_{s}r_{E}^{-\beta}}{N_{0}})^{2}}{}_{\phantom{}2}F_{1}\left(2,\frac{1}{2},\frac{5}{2},\frac{z\frac{P_{s}r_{E}^{-\beta}}{N_{0}}-1}{z\frac{P_{s}r_{E}^{-\beta}}{N_{0}}+1}\right).\label{eq:mgf-E-final}
\end{equation}

\begin{figure}[th]
\begin{centering}
\includegraphics[scale=0.45]{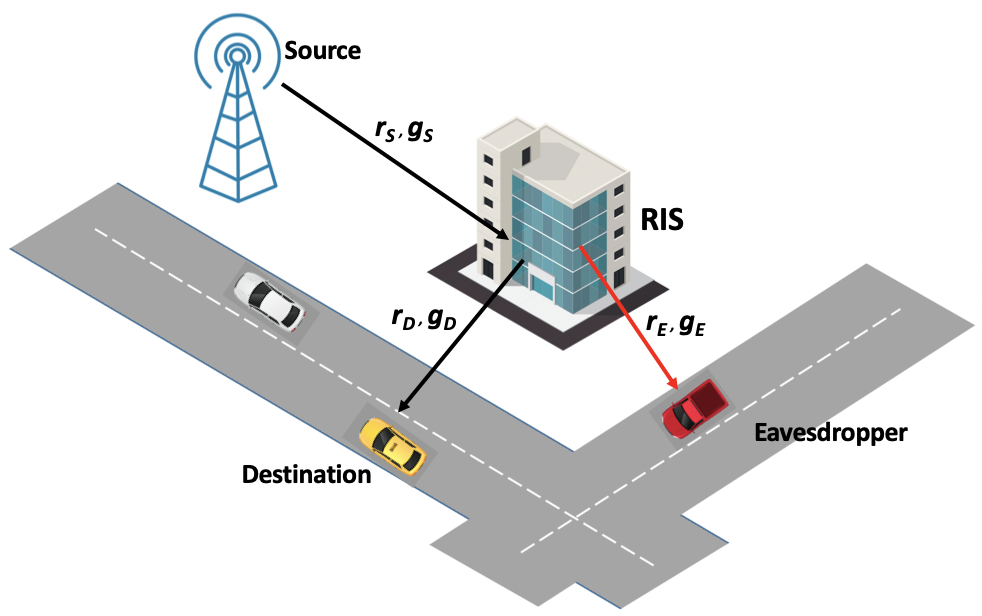}
\par\end{centering}
\caption{System model for a vehicular network scenario. Source station using
building-mounted-RIS as relay for vehicular communication.\label{fig:sys-mod-1}}
\end{figure}
From (\ref{eq:cs-defined-2}), (\ref{eq: log-id-2}), (\ref{eq:mgf-D-final})
- (\ref{eq:mgf-E-final}), the average secrecy capacity can be represented
as (\ref{eq:Cs-final}), shown at the top of the this page.

\begin{figure*}[t]
{\small{}
\begin{multline}
\overline{C}_{s}=\frac{1}{\textrm{ln}\left(2\right)}\stackrel[0]{\infty}{\int}\frac{1}{z}e^{-z}\left[\left(1-\left\{ \frac{4}{3\left(N_{0}+zP_{s}r_{D}^{-\beta}\right)^{2}}{}_{\phantom{}2}F_{1}\left(2,\frac{1}{2},\frac{5}{2},\frac{zP_{s}r_{D}^{-\beta}-N_{0}}{zP_{s}r_{D}^{-\beta}+N_{0}}\right)\right\} ^{N}\right)\right.\\
\qquad-\left.\left(1-\left\{ \frac{4}{3\left(N_{0}+zP_{s}r_{E}^{-\beta}\right)^{2}}{}_{\phantom{}2}F_{1}\left(2,\frac{1}{2},\frac{5}{2},\frac{zP_{s}r_{E}^{-\beta}-N_{0}}{zP_{s}r_{E}^{-\beta}+N_{0}}\right)\right\} ^{N}\right)\right]\textrm{d}z.\label{eq:Cs-final}
\end{multline}
}{\small\par}
\selectlanguage{american}%
\centering{}\rule[0.5ex]{2.03\columnwidth}{0.8pt}\selectlanguage{english}%
\end{figure*}

\section{VANET Transmission Through RIS Relay\label{sec:RIS-relay} }

In this section, we describe the VANET system transmitting through
a RIS relay and derive expressions for the average secrecy capacity
of the system.

\subsection{System Description}

In this section, we consider an RIS-based scheme with the RIS employed
as a relay or reflector for vehicular nodes in the network. Fig. \ref{fig:sys-mod-1}
illustrates the RIS-based system under consideration. The RIS is deployed
on a building and used as a relay for the signal from stationary source
$S$, while $D$ and $E$ are assumed to be highly mobile vehicular
nodes. Under this assumption, the source-to-RIS channel $g_{s}$ is
assumed to be Rayleigh faded, while the RIS-to-destination and RIS-to-eavesdropper
fading channels, $g_{D}$ and $g_{E}$ are assumed to be double-Rayleigh
distributed. The RIS is in the form of a reflect-array comprising
N reconfigurable reflector elements, capable of being controlled by
a communication oriented software for intelligent transmission. With
this in mind, the received signals at $D$ and $E$ are
\begin{equation}
y_{i}=\left[\sum_{n=1}^{N}h_{s}h_{i,n}e^{-j\phi_{n}}\right]x+w_{i},\label{eq:y}
\end{equation}
where $h_{s,n}=\sqrt{g_{s,n}r_{s}^{-\beta}e^{-j\theta_{n}}}$ is the
source-to-RIS channel with distance $r_{s}$, phase component $\theta_{n}$
and $g_{s}$ following a Rayleigh fading distribution. The term $h_{i,n}=\sqrt{g_{i,n}r_{i}^{-\beta}e^{-j\psi_{n}}},\quad i\in\left\{ D,E\right\} $,
is the channel coefficient from RIS-to-vehicle node, with distance
$r_{i}$, path-loss exponent $\beta$, phase component $\psi_{n}$
and $g_{i,n}$ following a double-Rayleigh distribution to model the
mobility of the nodes. \cite{psl_v2v}. The instanstaneous SNRs at
$D$ and $E$ are given respectively by
\begin{equation}
\gamma_{D}^{r}=\frac{\sum_{n=1}^{N}P_{s}\mid h_{s,n}\mid^{2}\mid h_{D,n}\mid^{2}}{N_{0}}\label{eq:snr-D-r}
\end{equation}
and
\begin{equation}
\gamma_{E}^{r}=\frac{\sum_{n=1}^{N}P_{s}\mid h_{s,n}\mid^{2}\mid h_{E,n}\mid^{2}}{N_{0}}\label{eq:snr-E-r}
\end{equation}

\subsection{Average Secrecy Capacity Analysis}

From (\ref{eq: log-id-2}), we obtain the average capacity for the
destination V2V link as 
\begin{equation}
\overline{C}_{D,r}=\frac{1}{\textrm{ln}\left(2\right)}\stackrel[0]{\infty}{\int}\frac{1}{z}\left(1-\mathcal{M}_{D,r}\left(z\right)\right)e^{-z}\textrm{d}z,\label{eq: log-id-3}
\end{equation}
where $\mathcal{M}_{D,r}\left(z\right)=\mathbb{E}\left[e^{-z\frac{P_{s}r_{s}^{-\beta}r_{D}^{-\beta}}{N_{0}}\sum_{n=1}^{N}g_{s,n}g_{D,n}}\right]$
is the MGF of the SNR at $D$. As for the joint distribution of $g_{s}$
and $g_{D}$, given that $g_{s}$ is a Rayleigh RV and $g_{D}$ is
a double-Rayleigh RV, we can define the RV $g=g_{s}g_{D}$, which
follows the cascaded Rayleigh distribution with $n=3$. From the generalized
cascaded Rayleigh distribution \cite{cascaded_ray}, we can evaluate
the PDF of $g$ as

\begin{equation}
f\left(g\right)=\frac{1}{\sqrt{2}}\textrm{G}_{0,3}^{3,0}\left(\frac{1}{8}g^{2}\Biggl|\negthickspace\begin{array}{c}
-\\
\frac{1}{2},\negthickspace\frac{1}{2},\negthickspace\frac{1}{2}
\end{array}\right).\label{eq: triple rayleigh pdf}
\end{equation}

Thus, the $\mathcal{M}_{D,r}\left(z\right)$ is given by
\begin{align}
\mathcal{M}_{D,r}\left(z\right)= & \mathbb{E}\left[e^{-z\frac{P_{s}r_{s}^{-\beta}r_{D}^{-\beta}}{N_{0}}\sum_{n=1}^{N}g_{s,n}g_{D,n}}\right]\nonumber \\
= & \prod_{n=1}^{N}\stackrel[0]{\infty}{\int}e^{-zg\frac{P_{s}r_{s}^{-\beta}r_{D}^{-\beta}}{N_{0}}}f_{g}(g)\textrm{d}g\nonumber \\
= & \prod_{n=1}^{N}\stackrel[0]{\infty}{\int}\frac{1}{\sqrt{2}}e^{-zg\frac{P_{s}r_{s}^{-\beta}r_{D}^{-\beta}}{N_{0}}}\textrm{G}_{0,3}^{3,0}\left(\frac{1}{8}g^{2}\Biggl|\negthickspace\begin{array}{c}
-\\
\frac{1}{2},\negthickspace\frac{1}{2},\negthickspace\frac{1}{2}
\end{array}\right)\textrm{d}g.\label{eq:mgf-D-1-1}
\end{align}

Using (\ref{eq: triple rayleigh pdf}) and \cite[Eq. (7.813.2)]{book2}
along with some basic algebraic manipulations, we can obtain the MGF
as 
\begin{equation}
\mathcal{M}_{D,r}\left(z\right)=\prod_{n=1}^{N}\frac{1}{z\mu_{d}\sqrt{2\pi}}\textrm{G}_{2,3}^{3,2}\left(\frac{1}{2\left(z\mu_{d}\right)^{2}}\Biggl|\negthickspace\begin{array}{c}
0,\frac{1}{2}\\
\frac{1}{2},\negthickspace\frac{1}{2},\negthickspace\frac{1}{2}
\end{array}\right),\label{eq:mgf-D-final-r}
\end{equation}
where $\mu_{d}=\frac{P_{s}r_{s}^{-\beta}r_{D}^{-\beta}}{N_{0}}$\textcolor{black}{.
Using similar analysis, the average capacity of the eavesdropper link
can be represented as}
\begin{equation}
\overline{C}_{E,r}=\frac{1}{\textrm{ln}\left(2\right)}\stackrel[0]{\infty}{\int}\frac{1}{z}\left(1-\mathcal{M}_{E,r}\left(z\right)\right)e^{-z}\textrm{d}z,\label{eq:av-CE-defined-r}
\end{equation}
where the MGF $\mathcal{M}_{E,r}\left(z\right)=\mathbb{E}\left[e^{-z\frac{P_{s}r_{s}^{-\beta}r_{E}^{-\beta}}{N_{0}}\sum_{n=1}^{N}g_{s,n}g_{E,n}}\right]$
and can be similarly evaluated as 
\begin{equation}
\mathcal{M}_{E,r}\left(z\right)=\prod_{n=1}^{N}\frac{1}{z\mu_{e}\sqrt{2\pi}}\textrm{G}_{2,3}^{3,2}\left(\frac{1}{2\left(z\mu_{e}\right)^{2}}\Biggl|\negthickspace\begin{array}{c}
0,\frac{1}{2}\\
\frac{1}{2},\negthickspace\frac{1}{2},\negthickspace\frac{1}{2}
\end{array}\right),\label{eq:mgf-E-final-1}
\end{equation}
where $\mu_{e}=\frac{P_{s}r_{s}^{-\beta}r_{e}^{-\beta}}{N_{0}}$.
\begin{figure}[th]
\begin{centering}
\includegraphics[scale=0.45]{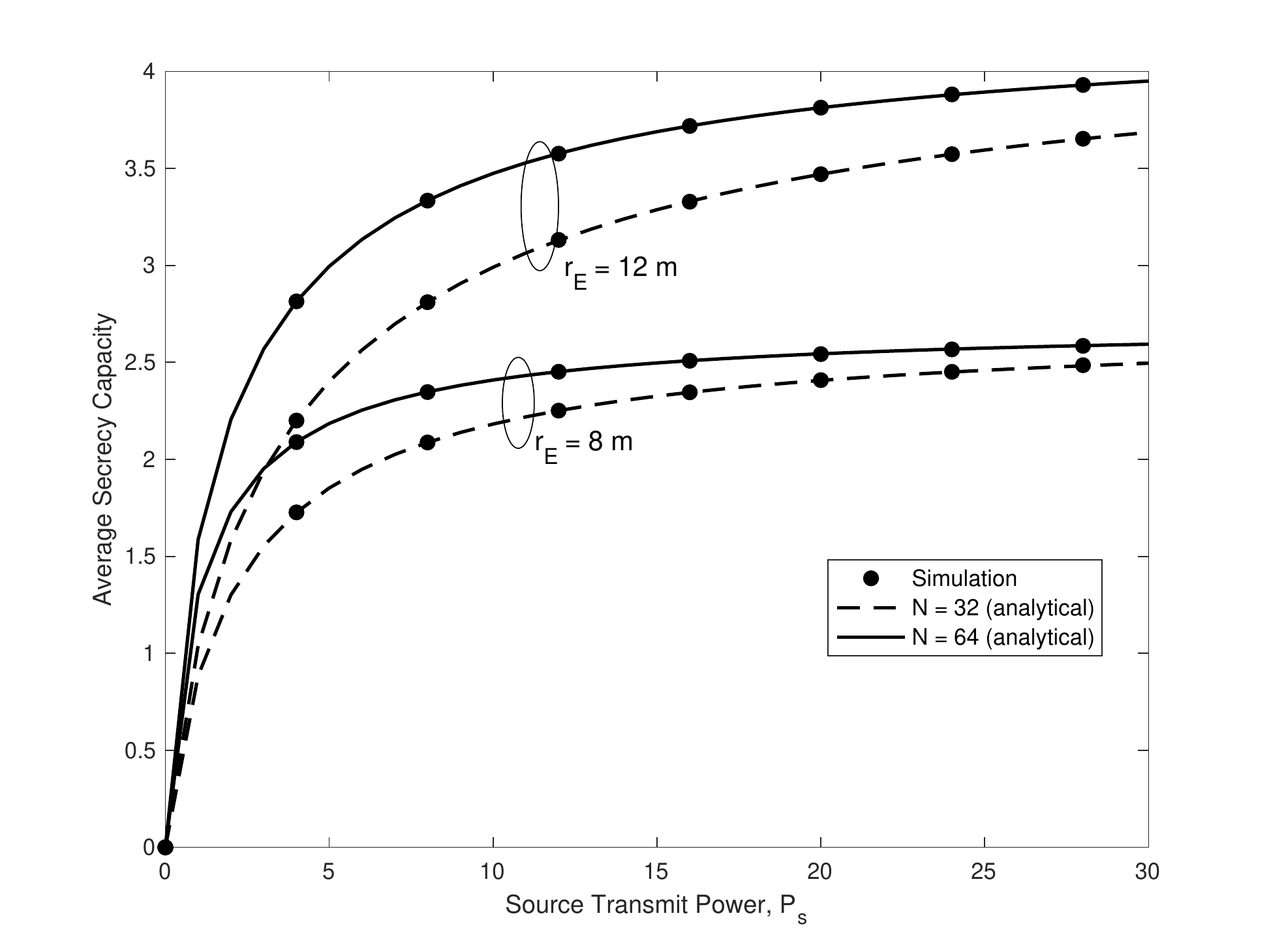}
\par\end{centering}
\caption{Average secrecy capacity versus source transmit power $P_{s}$ for
the V2V network with RIS as AP. Parameters considered with varying
eavesdropper distance $r_{E}$ and number of RIS cells $N$. \label{fig:cs-vs-ps-v2v}}
\end{figure}

From (\ref{eq:cs-defined-2}), (\ref{eq: log-id-3}), (\ref{eq:mgf-D-final-r})
- (\ref{eq:mgf-E-final-1}), the average secrecy capacity can be represented
as (\ref{eq:Cs-final-r}), shown at the top of the next page.

\begin{figure*}[t]
{\small{}
\begin{multline}
\overline{C}_{s,r}=\frac{1}{\textrm{ln}\left(2\right)}\stackrel[0]{\infty}{\int}\frac{1}{z}e^{-z}\left[\left(1-\left\{ \frac{N_{0}}{\sqrt{2\pi}zP_{s}r_{s}^{-\beta}r_{D}^{-\beta}}\textrm{G}_{2,3}^{3,2}\left(\frac{1}{2}\left(\frac{N_{0}}{zP_{s}r_{s}^{-\beta}r_{D}^{-\beta}}\right)^{2}\Biggl|\negthickspace\begin{array}{c}
0,\frac{1}{2}\\
\frac{1}{2},\negthickspace\frac{1}{2},\negthickspace\frac{1}{2}
\end{array}\right)\right\} ^{N}\right)\right.\\
\qquad-\left.\left(1-\left\{ \frac{N_{0}}{zP_{s}r_{s}^{-\beta}r_{E}^{-\beta}\sqrt{2\pi}}\textrm{G}_{2,3}^{3,2}\left(\frac{1}{2}\left(\frac{N_{0}}{zP_{s}r_{s}^{-\beta}r_{E}^{-\beta}}\right)^{2}\Biggl|\negthickspace\begin{array}{c}
0,\frac{1}{2}\\
\frac{1}{2},\negthickspace\frac{1}{2},\negthickspace\frac{1}{2}
\end{array}\right)\right\} ^{N}\right)\right]\textrm{d}z.\label{eq:Cs-final-r}
\end{multline}
}{\small\par}
\selectlanguage{american}%
\centering{}\rule[0.5ex]{2.03\columnwidth}{0.8pt}\selectlanguage{english}%
\end{figure*}

\section{\textcolor{black}{Numerical Results and Discussions\label{sec:Results}}}

\textcolor{black}{}
\begin{figure}[th]
\begin{centering}
\textcolor{black}{\includegraphics[scale=0.45]{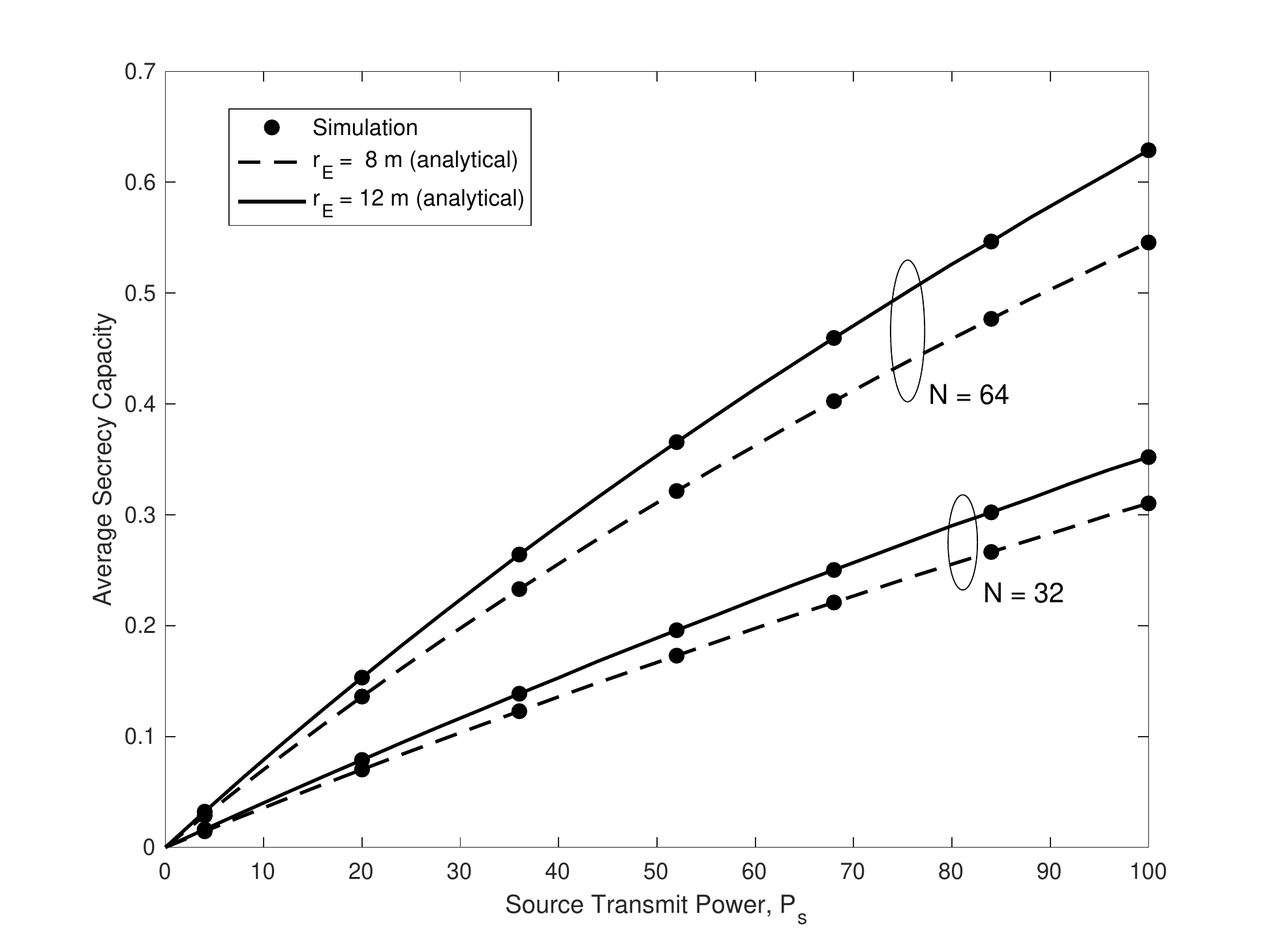}}
\par\end{centering}
\textcolor{black}{\caption{Average secrecy capacity versus source transmit power $P_{s}$ for
the VANET with RIS as relay. Parameters considered with varying eavesdropper
distance $r_{E}$ and number of RIS cells $N$.\label{fig:cs-vs-ps-relay}}
}
\end{figure}

\textcolor{black}{In this section, we present and discuss some results
from the mathematical expressions derived in the paper. We then investigate
the effect of key parameters on the secrecy capacity of the system.
The results are then verified using Monte Carlo simulations with at
least $10^{6}$ iterations. Unless otherwise stated, we have assumed
}source power $P_{s}=10$W, RIS-to-$D$ distance $r_{D}=4$m, RIS-to-$E$
distance $r_{E}=8$m, source-to-RIS distance $r_{s}=10$m and pathloss
exponent $\beta=2.7$.

In Fig. \ref{fig:cs-vs-ps-v2v}, we commence analysis for the V2V
network model, with the average secrecy capacity against source power
for different numbers of RIS cells and eavesdropper distances. It
can be observed that the secrecy capacity increases with an increase
in $P_{s}$, $r_{E}$ or $N$. It can be further noted that within
the region considered, the eavesdropper distance has a greater effect
on the secrecy capacity than doubling the number of RIS cells. Also,
the effect of increased RIS cells, is more pronounced when the eavesdropper
is further away. A similar analysis can be made for the average secrecy
capacity of the VANET RIS-relay model considered, as observed in Fig.
\ref{fig:cs-vs-ps-relay}. However, the effect of increased source
power produces an almost linear response for the secrecy capacity,
while the effective value of the secrecy capacity is much lower than
the V2V RIS model for similar $P_{s}$.

\begin{figure}[th]
\begin{centering}
\textcolor{black}{\includegraphics[scale=0.45]{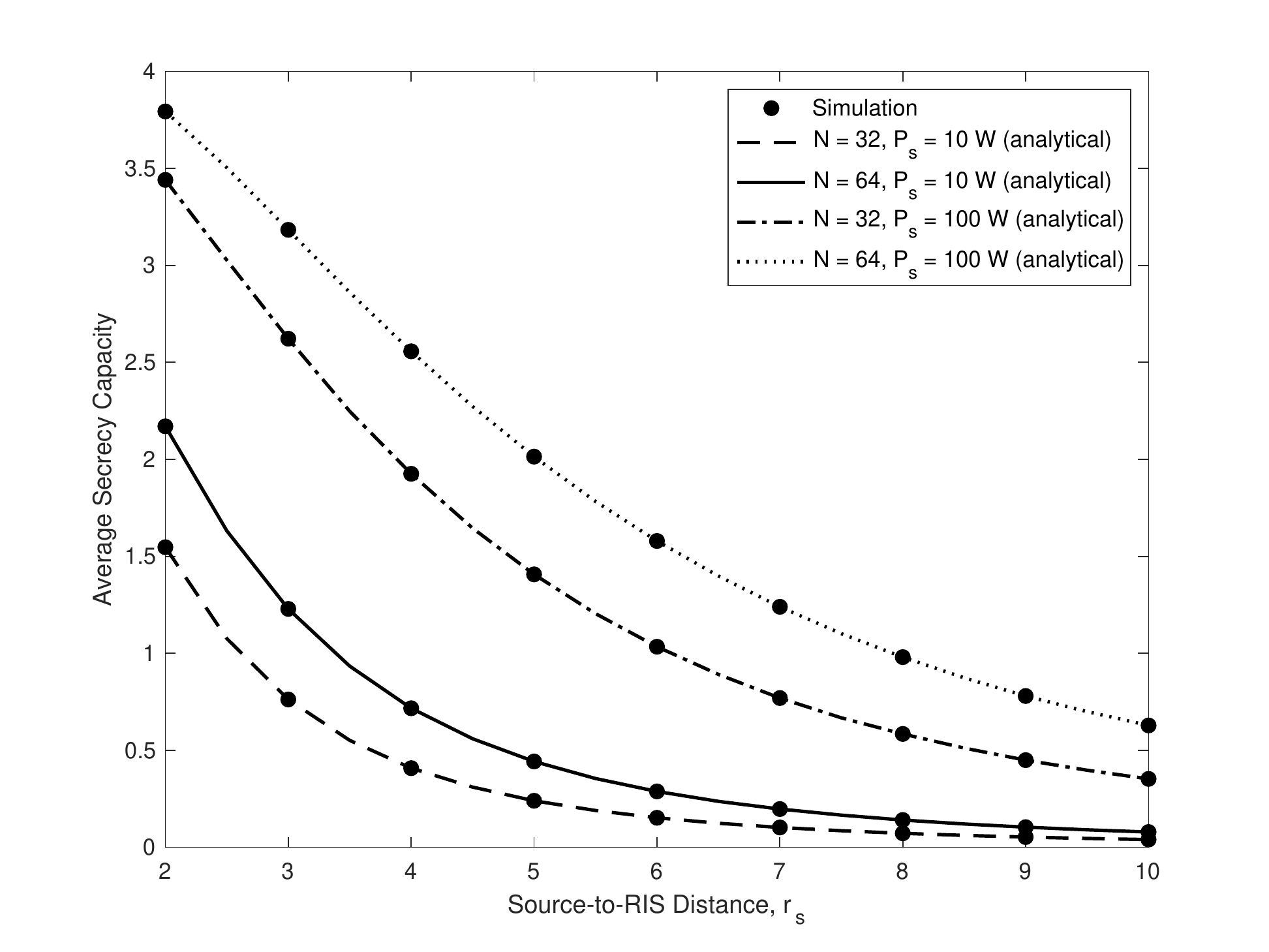}}
\par\end{centering}
\textcolor{black}{\caption{Average secrecy capacity versus source-to-RIS distance for the VANET
with RIS as relay. Parameters considered with varying source transmit
power $P_{s}$ and number of RIS cells $N$.\label{fig:cs-vs-rs-relay}}
}
\end{figure}

Fig. \ref{fig:cs-vs-rs-relay}, shows a plot of the average secrecy
capacity against $r_{s}$ with different values of $P_{s}$ and $N$,
for the VANET RIS-relay system. We assume the RIS-to-eavesdropper
distance $r_{E}=12$m. First, we observe that the average secrecy
capacity decreases as the source distance increases, demonstrating
the effect of fading and pathloss on the link, before the RIS relay.
The result also demonstrates that doubling the number of RIS cells
has less influence on the average secrecy capacity, as compared to
the impact of the source power, within the observed region.\textcolor{black}{}

\section{\textcolor{black}{Conclusions\label{sec:Conclusions}}}

\textcolor{black}{In this paper, we examined the secrecy capacity
}as a key metric for PLS of a wireless vehicular communication network.
Two scenarios of a RIS-based vehicular network were considered. The
results demonstrate how the secrecy capacity of a vehicular network
can be improved with respect to the source power, eavesdropper distance
and the number of RIS cells. The results further showed how the location
and size of RIS (in terms of number of RIS cells) can be employed
to improve a RIS relay-based VANET. 

\bibliographystyle{IEEEtran}
\bibliography{bibGC19}

\end{document}